# A SURVEY OF USER-CENTRIC DATA WAREHOUSES: FROM PERSONALIZATION TO RECOMMENDATION

Eya Ben Ahmed[1], Ahlem Nabli[2] and Faïez Gargouri[3]

[1]University of Tunis, High Institute of Management of Tunis, Tunisia
eya.benahmed@gmail.com
[2] University of Sfax, Faculty of Sciences of Sfax, Tunisia
ahlem.nabli@fsegs.com
[3]High Institute of Computer and Multimedia of Sfax, Tunisia
Faiez.gargouri@isimsf.com

## ABSTRACT

*Providing a customized support for the OLAP brings tremendous challenges to the OLAP technology. Standing at the crossroads of the preferences and the data warehouse, two emerging trends are pointed out; namely: (i) **the personalization** and (ii) **the recommendation**.*

*Although the panoply of the proposed approaches, the user-centric data warehouse community issues have not been addressed yet.*

*In this paper we draw an overview of several user centric data warehouse proposals. We also discuss the two promising concepts in this issue, namely, the personalization and the recommendation of the data warehouses. We compare the current approaches among each others with respect to some criteria.*

## 1. INTRODUCTION

Multidimensional databases (MDB) are large collections of historical data, frequently used by decision makers for complicated analysis and complex tasks. Within these databases, data are organized according to subjects of analysis, called facts, and axes of analysis, called dimensions. Dimensions are usually organized as hierarchies, supporting different levels of data aggregation.

By offering multidimensional data views and the pre-computation of summarized data, data warehouse systems are well suited for on-line analytical processing (OLAP). OLAP analyses are performed through interactive exploration of MDB through a set of navigational operators like *drilldown*, *rollup*, *slice* and *dice*. Such operations accommodate several user viewpoints. Nevertheless, the MDB exploration problems arise when data dimensionality increases. Therefore, user analyses become effortful, complicated and time consuming.

OLAP analysis provides an interactive investigation. In general, such task suffers from several deficiencies. In fact, facing an analysis query, several outcomes of this query are plausible: first, an empty result can be returned to the user. Second, information flooding can be generated. In addition, inappropriateness of the user expectations for the obtained results is majorly possible. Furthermore, a heavy-tailed runtime due to the generality of launched queries can be consumed.

To avoid such fallacies, extensive efforts have been devoted to increasing the user implication on the exploration task. In fact, two streams of approaches for OLAP database systems extension emerged, namely:

(i). Approaches privileging *implicit* intervention through the customizability of the user behavior according to their preferences. Such strategy is called **personalization** [4,5,6,8,9,15,18,19,21,22,26,27,28];

(ii). Approaches favoring *explicit* intervention through applying **recommendation** to better assist the user on his decision making even he does not accurately discern the data warehouse schema [1,2,7,16,17].

In the sequel, we closely scrutinize the personalization and the recommendation concepts [23]. Then, we draw an overview of the proposed approaches in each trend. Finally, we present a comparative study confronting the panoply of the proposed approaches.

The remainder of the paper is organized as follows: Section 2 sketches the personalization and the recommendation. We present in section 3 a thorough study of the existing approaches in the user centric data warehouse area. We relate in section 4 a comparative study between the above mentioned contributions. Finally, section 5 concludes our work and presents some insights for future work.

## 2. PERSONALIZATION VS RECOMMENDATION

Two main strategies emerged in the user centric data warehouse: (i) *personalization*, (ii) *recommendation*. In this section, we exhaustively scrutinize them.

### 2.1. Personalization

The personalization is defined as a mechanism providing an overall customized, individualized user experience by taking into account the needs, preferences and characteristics of a user or group of users [11,12,13,14]. Generally, personalizing a system consists in defining and exploiting a user profile [3].

In OLAP context, the most emerging axis is the query personalization. The main idea behind its process consists in considering the user preferences when answering query [6,8,15,18,19, 21,22,26,27]. In fact, the personalization of the query is considered as an addition of selection conditions extracted from the user profile. Therefore, different users may find and see the facts they prefer without any loss of time and effort on navigation. Thus, they may obtain distinct responses to the similar queries, based on personal preferences stored on their profiles.

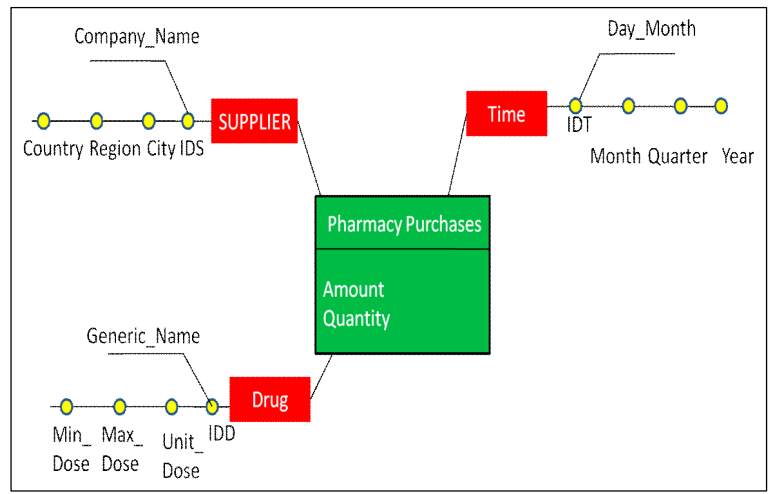

Figure 1. Star schema of pharmacy purchases

To illustrate this concept, we present in Figure 1 an example of a constellation that allows analysing online purchases in a pharmacy.

*For instance, we consider two decision-makers covering different roles and sharing the same data warehouse (cf. Figure 1).*

*The **purchases manager** in the pharmacy is basically interested in monthly purchases, but he may also wish to see more detailed data if the weekly purchase exceeds a given amount. However **a purchase agent** working in a given region prefers to see highest commissions data in that region.*

*Having different interests and several objectives on analysing the dataset, different users need a tiring effort and wasted time in queries formulation to achieve their goals.*

*The two decision markers can obtain different results to the same OLAP query using their preferences through the personalization of the data warehouse.*

### 2.2. Recommendation

The second technique consists in recommendation. The recommender Systems (RS), for instance, automatically suggest new content that should comply with the user's taste. They help users navigating through large product assortments and making decisions.

Probably the most prominent example is the book recommendation service of the e-tailer Amazon.com. These personalized services share a common concern: modeling the user's taste. Therefore, such systems need to somehow capture likes and dislikes in order to model or infer the user's preferences. In fact, several data mining algorithms were employed to design the RS.

In OLAP context, the key idea of recommendation is to leverage what the other users did during their former navigations on the cube or what the user explored during the previous sessions, and use this information as a basis for recommending to the user his forthcoming query [1,2,16,17]. Both partial and complete queries can be provided to users [16,17].

*For instance, we consider the same data warehouse (cf. Figure 1), the **purchases manager** can forget on expressing his analysis queries some interesting attributes which may be fundamental for efficient responses.*

*For example, he analyses the monthly purchases, but omits to focus on the most bought drug. So, the system can provide recommendation to usefully propose him this possibility.*

*Or the system can also offers more detailed data if the weekly purchases of given product are outstandingly increasing in the respect of a given growth rate.*

*Another alternative is to anticipate the purchases manager analysis and to propose customized recommendations to efficient orient him on his analysis process.*

Noticeably different of the personalization, the process of recommendation from an initial query outputs a recommended query obviously different from the initial one due to disjoint user interests.

## 3. CLASSIFICATION OF USER-CENTRIC DATA WAREHOUSE APPROACHES

In this section, we present a thorough survey of approaches dealing with the user centric data warehouses. This challenge seemed to be a promising issue.

According to the proactiveness aspect, two main pools can be distinguished: (i) **personalization** approaches that reformulate the user queries based on his preferences and (ii) **recommendation** approaches that propose new queries based on the navigation log or based on the analysis context.

### 3.1 Personalization

In the respect of the subject of the personalization, existing data warehouse personalization approaches are usually classified into two categories**:**

(i) *Content-based warehouse personalization approaches* [4,5,6,8,9,15,18,24,26,27,28,29] usually are based on the content, particularly on customizing the data warehouse schema through the user preferences and needs.
(ii) *Hybrid warehouse personalization approaches* [19,21,22] combine the content as well as the form personalization. Indeed, not only the schema of the data warehouse is adapted to the user choices but also the visualization of analysis is tailored according to the user interest.

#### 3.1.1 Contentóbased warehouse personalization

Ravat et al. [8,9] focus on the integration of personalization in a multidimensional context. The proposed conceptual model is based on multidimensional concepts (fact, dimension, hierarchy, measure, parameter or attribute). To assign priority weights to attributes of a multidimensional schema, the personalization rules are described using the *Condition-Action* formalism. Accordingly, an OLAP query language [20] adapted to the personalization context is proposed. The weights are taken into account during OLAP analyses. For instance, displaying data with higher weights can outstandingly reduce the number of eventual multidimensional operations that the user can launch. Or, fixing a minimum threshold may induce that only attributes whose weights exceed this threshold will be displayed, the remainder attributes should be explicitly mentioned to display them.

In addition, the proposed algebra contains OLAP operators allowing the drilling, rotations, selection, ordering, aggregation and modification operations.

For instance, we present the syntax of the rule definition command associated to either fact (NC), either one-dimensional (ND) either a hierarchy ($N^{hD}$ i) with an illustrative example used to personalize the constellation schema of the data warehouse (cf. figure 2).

| Rule Format | Example |
|---|---|
| **CREATE RULE ON** <RuleName> <$N^D$> | <$N^{hD}$ i> |<$N^F$><br>**WHEN** <manipulation><br>[**IF** <condition>]<br>**THEN** <action>;<br><br><**manipulation**> : *Determines the current context of manipulation defined using the launched operation (i.e., Displayed, Rotated, Drilled-down, Rolled-Up) on the components of the constellation.*<br><**Condition**>: *concerns the condition related to the current state of the constellation.*<br><**Action**>: *measures or dimensions of the constellation applied to the facts.* | **CREATE RULE** IDS_Rule **ON** SUPPLIER.IDS<br>**WHEN** DISPLAYED OR ROTATED<br>**IF** isCurrent('Sales')<br>**THEN**<br>  **BEGIN**<br>    setWeight('City', 0.5);<br>    setWeight('Region', 0.8);<br>    setWeight('Country', 1);<br>  **END**; |

Figure 2. Rule definition with an example

This solution has been implemented and outputs a personalized multidimensional database system allowing users both to define personalized rules and to query the personalized database.

Jerbi et al [15,18] propose a context-aware OLAP Preference model which is defined on MDB schema. Using a qualitative approach, the OLAP preferences are modeled and closely depend on user analysis context (c.f. figure 3). That's why a conceptual model of OLAP context is conceived using an arborescence of OLAP analysis elements.

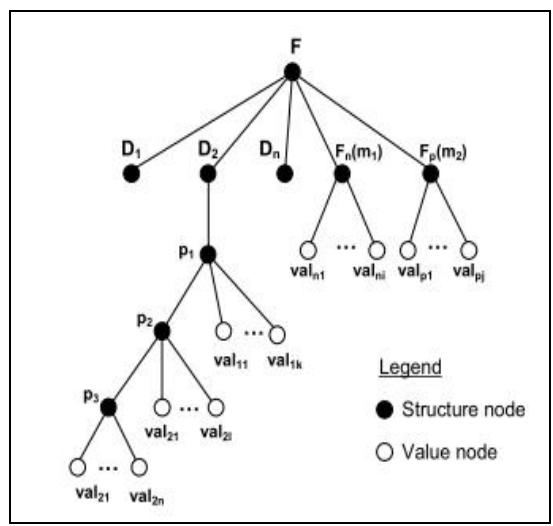

Figure 3. OLAP analysis context Tree

In addition, a query personalization framework is introduced. Its process of an initial query proceeds in two steps:
   (i). ***User preference selection*** where preferences related to the context query are extracted. Such preference location problem is considered as trees matching problem: a preference whose context tree is included in the analysis context tree is a candidate preference.
   (ii). ***Query enhancing*** where the related preferences are incorporated dynamically into the initial query. The next step is to enhance the original query with the maximal of elements according to the partial ordering of the selected preference. The augmented query is executed providing personalized aggregated data.

Rizzi and Golfarelli [24, 27,28] extend the proposal of Kiessling [29] to the OLAP domain to introduce MyOLAP approach where preferences are formulated by the user for each single query using a visual interface or using an extended version of MDX language. They are also considered as soft constraints tailored for the OLAP context.

Formulated on schema, the preferences concern not only dimensional attributes but also measures, and group-by sets. The preferences composition is modeled using predicate logic attributes and expressed through two main methods: (i) Pareto composition pareto when two preferences are equally relevant; (ii) Prioritization when a preference is more relevant than another.

To answer preference query on a data cube according to the best match only model (BMO), only the facts not worse than any other facts are returned using the WeSt algorithm. Its key idea is to create a new type of partitioning graph whose nodes collapse several classes into one node.

Favre et al [4,5,6] consider that the personalization of the data warehouse is an evolution of analysis needs in data warehouses. A formal model describing an evolving data warehouse is proposed. It is composed of "**if-then**" rules that support the personalization process. Those aggregation rules are composed of a "fixed" part which corresponds to the warehouse schema and reflects global analysis needs, and an "evolving" part which is defined by the aggregation rule creating new granularity levels on the dimension hierarchies, satisfying the individual analysis needs.

A typical example of an aggregation rule is:
$$r: \quad \textit{if Drug\_Generic\_Name='Correctol' then}$$
$$\textit{Drug\_Min\_Dose=2 AND Drug\_Max\_Dose = 6}$$

The versioning of rules is expressed by managing different versions of hierarchies. Such versioning is closely dependant on the user and absolutely neglects any consideration of temporal version. Besides, the creator of any new analysis dimension may annotate it aiming to give a good description for a better understanding by others users.

Such personalization process is supported by an architecture composed of four steps:
   (i) *Knowledge acquisition* gathers users' knowledge in the form of aggregation rules ;
   (ii) *Integration* of these aggregation rules in the data warehouse ;
   (iii) *Data warehouse schema evolution* that can create a new granularity level through enriching an existing dimension hierarchy existing, or creating a new one;
   (iv) *On-line analysis* based on the new schema.

The overall evolution process is implemented and generated the WEDriK plateform [4].

Thereafter, Khemiri and Bentayeb [26] proposed an innovative personalization approach. Its driving idea is to create a materialized view of data warehouse according to each user profile. The latter is composed of a set of flexible preferences which are "temporary" and a set of rigid preferences which are "permanent". The profile definition is carried out offline when the user specifies at first time his preferences. Facing the complexity of managing preferences, the concept of groups is implemented.

The created view contains rigid preferences. When a query is launched, the system uses the materialized view to effectively answer instead of consulting all data warehouse and integrates the flexible preferences.

It is noteworthy that it is promising for each user to have its own data warehouse because such accurate determination of a specific view for each user, allows getting rapidly answers. To validate this approach, this solution is implemented using the NetBeans environment and SQL Server 2005 as DBMS. User profiles are stored in XML.

It should be noted that the construction of views and updates for each user is a costly and complex. So the usability of this proposal is restricted to context with limited number of users.

### 3.1.2 Hybrid warehouse personalization: Form and content-based

Bellatreche et al. [21] proposed an approach for personalizing answers to OLAP queries and providing the best visualization to the user based on the user profile. The latter contains on one side the preferences that rank members of all dimensions and on the other side the visualization constraints that control the form of the visualization of the query results.

The preferences define a pre-ordering on cubes through associating the degrees of interest. In fact, a member or value of an attribute is preferred to another member if its degree of interest is higher.

To answer an OLAP query, the preferences are used to personalize user queries through enrichment of the initial query by added selection conditions. Then, the constraint visualization is explored to present the best response visualization. This means that the framework personalizes both the set of facts to be presented to the user, and the presentation of these facts.

It is notably that the proposed model is very simple model of cube, where each dimension consists of only one attribute. However, expressing exhaustively a total pre-ordering on members for user preferences deprives the model from flexibility.

Unlike the previous contribution, [22] propose a new supple model based on a partial ordering of cubes. Besides, the response to any query may contain measures with different levels of aggregation. In addition, several members of dimensions can be combined.

After that, [19] sketches two methods of personalization before and after the query execution.

## 3.2 Recommendation

According to the generation process, existing data warehouse recommendation approaches can be splited into the following categories:
   (i) *Content-based methods* [16,17] recommend to the user items similar to the ones the user preferred in the past relying on product features and textual item descriptions;
   (ii) *Collaborative Filtering* [1,2,7] recommends to the user items that people with similar preferences liked in the past taking the behavior, opinions and tastes of a large community of users into account.

### 3.2.1 Contentóbased warehouse recommendation

According to user specific details, Jerbi et al. [16,17] distinguishes three categories of flexible recommendations that can be generated:

(i) *Interactive support in Querying*: when the user can explore the multidimensional data using graphical visualization, such as dragging elements from a navigation zone or refining the view;

(ii) *Anticipatory Recommendations:* Aiming to considerably reduce the user effort and the time spent, Jerbi et al. anticipate the user recommendation;

(iii) *Alternatives Recommendation*: The alternatives include providing more detailed information, reminding the user omitted information, offering additional information which the user did not ask for however they are so pertinent.

The OLAP analyses are represented using a graph-based model. In fact, both of user profile and current query are expressed by means of trees. Then, a Tree Matching Algorithm is applied to compare the two trees. A preference whose context tree is included in (all its edges and nodes belong to) the tree of the current analysis context is a candidate preference. If there are several candidate preferences, the selected preference is the most relevant one. Finally, the resulting query obtained from the transformation of the original query tree is proposed as a recommendation to the user.

This method of recommendation of OLAP queries based on the content uses an heuristic for the extrapolation on assigning scores to the unweighted nodes in the tree of preferences.

The proposed OLAP recommender system allows users to perform full or partial queries and to ask for help to build their analysis report.

### 3.2.2 Collaborative filtering warehouse recommendation

To apply a collaborative work approach that leverages former explorations of the cube to recommend OLAP queries, Giacometti et al [1,2] and Negre [7] proposed to explore the OLAP query log to assist the user on his analysis. Therefore, a framework for summarizing OLAP query logs is introduced. The major idea is that a query can summarize another query and that a log can summarize another log.

First, they proposed a query manipulation language (called QML) tailored for OLAP queries and composed of binary and unary operators that allow summarizing queries.

Second, to measure the best extension of a query is considered as a good summary of another query, a quality measure based on the classical notions of precision and recall is used.

Finally, a greedy algorithm for automatically constructing, using QML, a summary of a query log i.e., a quality threshold given by the user is proposed.

After summarizing a log of OLAP queries, Negre proposed a new framework for recommending such OLAP queries. The framework uses the query log of the OLAP server and the current session to recommend the next query. The three-phase based approach consists in:
- (i). Use of the query set partitioning to partition the query log in order to compute all the generalized sessions of the log which are groups of similar queries;
- (ii). Generating candidate recommendations by first finding which sessions of the log match the current session and then predicting what the forthcoming query can be. The used algorithm for computing recommendation is based on three functions, namely, Match, Predict, ClassRep with Match function is used to Łnd a set of generalized sessions matching a given generalized session and Predict is used to compute, for a set of generalized sessions, a set of candidate classes and ClassRep is used to obtain the query that represents a class ;
- (iii). Ranking the candidate recommendation so as to present to the user the most relevant queries first.

However, we shed light on the recommendations provided irrespective of user preferences while such preferences play an important role in the success of recommender systems.

Besides, this approach consists in recommending full queries and does not consider flexible recommendations that deal with different levels of user involvement.

## 4. DISCUSSION

Table 1 reports noteworthy a comparison of the above approaches according to several criteria.

Generally, the **proactiveness** characteristic distinguishes the *personalization* approaches and the *recommendation* approaches. In fact, only Jerbi et al. [16,17], Giacometti et al. [1,2] and Negre [7] are interested on providing explicitly recommendation to support the decision maker. Bellatreche et al. [21,22], Mouloudi[19], Ravat and Teste [8,9], Jerbi et al.[15,18], Rizzi [27,28], Favre et al. [4,5,6] and Khemiri and Bentayeb [26] focus on the personalization of the user queries to bring more flexibility to the system and to adapt the responses to his preferences.

In addition, two **types of customization** are pointed out: a *qualitative* (ordinal) formulation or *quantitative* expression. Due to the limited expressive power of the quantitative approaches, Jerbi et al [15,16,17,18] and Negre [7] expressed the qualitative preferences model using binary preference relations. The rest of the approaches dealing with quantitative formulation associate a numerical score to every tuple of the query answer.

Such customization has a major core outlining the **subject**. Indeed, the personalization or the recommendation is based on to the *content* or the *combination* of the content and the form or to the *collaborative filtering* which relies on the similarity of users preferences to predict the future preferences. Indeed, [4,5,6,8,9,15,18,26,27,28] focus on the content-based personalization, particularly the adaptation of the data warehouse schema to the user needs. Bellatreche et al. [21,22] and Mouloudi [19] combined the content personalization and the form visualization to better response to the userøs needs. However, Jerbi et al. [16,17] tackle the content-based recommendation inspired from the previous preferences of the user. According to the preferences of the similar users, Negre opted for the collaborative filtering recommendation. Nevertheless, the combination of content-based and the collaboration filtering sill not affordable.

Such customization is expressed using **operators** determinating how the strategy of personalization or recommendation is achieved using *a language preference* or a *method of customization*. Distinctly [8,9] and [27,28] proposed a new language to manage the user preferences. However, all other approaches dealing with personalization as well as recommendation use method to adapt the system to the user preferences.

The reported approaches have a limited **scope of customization** which presents the elements that may be redefined to response to the userøs needs, namely the *queries* or the *operations;*
All the proposed approaches rely on a redefinition of the analysis queries through adding constraints to restrict the OLAP answers. However, Ravat and Teste [8,9] and Jerbi el al.[15,18] proposed to redefine the OLAP operations, namely, the drilling, the rotations, the selection, the ordering, the aggregation and the modification operations.

Certainty, several works addresses the **source of preference elicitation** issue. Indeed, both the behavior and the analysis context can be considered as sources to generate the user preferences. Based on the *context*-aware preferences, Ravat and Teste [8,9] and Jerbi el al. [15,16,17,18] proposed both personalization and recommendation. All other approaches concentrate on the *behavior* of the current user or the users with similar preferences.

After elicitation the preferences, two trends of **preference formulation** emerged. In fact, the majority of approaches involve the user to *manually* specify his preferences, except the contribution of Negre et al [7] where the profile is built *automatically*.

Those preferences are used to query the data warehouse. The **timing of personalization or recommendation** is an important classification criteria. It presents the moment when the personalization or the recommendation is accomplished. It can be either *before* or *after querying* the data warehouse. Differently from the other proposals, Mouloudi [19] introduce the personalization of results after querying the data warehouse.

Regarding the **constraints**, two cases are possible: the proposal can basically neglect *any constraint* or can take *machinal constraints* into account. To better visualize the OLAP responses, Bellatreche et al. [21,22] and Mouloudi [19] integrated the visualization constraints such as the structure of the cross-table used to display the results.

To implement the above approaches, many types **of storage of data warehouses** are used in the approaches implementation, namely, (i) *Relational OLAP (ROLAP)* where the data warehouse are stored in a database relational data, (ii) *Multidimensional OLAP (MOLAP)* where data is stored in a multidimensional or (iii) *Hybrid OLAP (HOLAP)* where data are stored in a relational database, and aggregations are stored in multidimensional structures.
Bellatreche et al.[21,22], Mouloudi [19], Favre et al [4,5,6] and Khemiri and Bentayeb [26] implemented their contribution using the MOLAP. However, the remainder approaches are implemented using a ROLAP.

Technically, these proposals have their roots in different **inspiration fields** such as the *information retrieval (IR),* the *database (DB)* [10,25] and *Human Computer Machine (HCM).* In fact, all the propositions are inspired from the database personalization domain except the contribution of Rizzi [27,28] who uses the HIM personalization to present this proposal. Besides, both of Bellatreche et al. [21,22] and Mouloudi [19] use the IR domain to build the user profiles.

| Proposal | | Bellatre-che et al. (2005-2006) | Mou-loudi (2007) | Ravat and Teste (2007-2008) | Jerbi et al. (2008, 2010) | Jerbi et al. (2009) | Giacometti et al Negre (2008-2011) | Rizzi (2007-2010) | Favre et al. (2007) | Khemiri and bentayeb (2010) |
|---|---|---|---|---|---|---|---|---|---|---|
| Proactivess | Personalization | X | X | X | X | | | X | X | X |
| | Recommendation | | | | | X | X | | | |
| Type | Qualitative | | | | X | | | X | | |
| | Quantitative | X | X | X | | X | X | | X | X |
| Subject | Content | X | X | X | X | X | | X | X | X |
| | Form | X | X | | | | | | | |
| | Collaborative | | | | | | X | | | |
| Operator | Language | | | X | | | | X | | |
| | Method | X | X | | X | X | X | | X | X |
| Scope | Query | X | X | X | X | X | X | X | X | X |
| | Operation | | | X | X | | | | | |
| Source | Behavior | X | X | X | | X | X | X | X | X |
| | Context | | | | X | | | | | |
| Preference Formulatio | Manuel | X | X | X | X | X | | X | X | X |
| | Automatic | | | | | | X | | | |
| Time | Before quering | X | X | X | X | X | X | X | X | X |
| | After quering | | X | | | | | | | |
| Constraint | No constraint | | | X | | X | X | X | X | X |
| | Machinal | X | X | | X | | | | | |
| Storage | ROLAP | | | X | X | X | X | X | | |
| | MOLAP | X | X | | | | | | X | X |
| | HOLAP | | | | | | | | | |
| Inspiration field | HCM | | | | | | | X | | |
| | IR | X | X | | | | | | | |
| | Database | X | X | X | X | X | X | | X | X |

Table 1. Comparison of surveyed approaches dealing with user-centric data warehouse

## 5. CONCLUSION

In this paper, we first draw an overview of user-centric data warehouse **customizability**. Current approaches were confronted according to several criteria.

In fact, the studied issues allow us introduce the core of our future work which the proposal of a new approach on data warehouse personalization. The purpose of this work is to take advantages of the studied contributions in an optimized way.

## ACKNOWLEDGEMENTS

The authors would like to thank everyone!